# SCIENTOMETRICS AND COMMUNICATION THEORY:
# TOWARDS THEORETICALLY INFORMED INDICATORS

Loet Leydesdorff
*Science and Technology Dynamics*
Amsterdam School of Communications Research (ASCoR)
Kloveniersburgwal 48, 1012 CX  Amsterdam, The Netherlands
&

Peter Van den Besselaar
Netherlands Institute for Scientific Information Services (NIWI)
Postbus 95110, 1090 HC  Amsterdam, The Netherlands

**Abstract**

A theory of citations should not consider cited and/or citing agents as its sole subject of study.  One is able to study also the dynamics in the networks of communications.  While communicating agents (e.g., authors, laboratories, journals) can be made comparable in terms of their publication and citation counts, one would expect the communication networks not to be homogeneous.  The latent structures of the network indicate different codifications that span a space of possible "translations".  The various subdynamics can be hypothesized from an evolutionary perspective.  Using the network of aggregated journal-journal citations in *Science & Technology Studies* as an empirical case, the operation of such subdynamics can be demonstrated.  Policy implications and the consequences for a theory-driven type of scientometrics will be elaborated.

**Introduction**

A pervasive property of scientometric distributions is their skewness.  This property was noted early in the history of the discipline (e.g., Lotka[1]), and it has been used as a starting point for theorizing.  Price,[2] for example, suggested that skewness could provide a basis for "a general theory of bibliometric and other cumulative advantage processes."  The system of reference for this theory, however, remained the communicating agent(s).

The skewness of the distribution can be appreciated also in terms of the selectivity of the relevant communication networks.  The communication systems produce the skewness in the distributions since they are based on (recursive) selections upon selections by the carrying agents. The power-laws which describe these skewed distributions can then be considered as an indicator of self-organization in the emerging systems of communication (Katz, *personal communication*; cf. Bak & Chen[3]). The emerging order is not an intended outcome of conscious agency, but a result of the interactions among agents.

If one models the actors and their relations as the row-vectors of a matrix, the communication systems are represented by the columns. Thus, networks of communication are structurally coupled to the carrying agents, but they develop in orthogonal dimensions and according to different dynamics (cf. Maturana[4]). In general, the selectivity in the communication leads to sparse matrices, and thus skewness in the distribution. The latter is reflected in the so-called "propensity to cluster" in scientometric analyses of such matrices. Precisely because of this pervasive property of distributions that reflect highly specific communications, scientometric indicators can sometimes be robust. The structure is so pronounced that it cannot be suppressed in the representation, almost independently of the type of statistics used.

Scientometric mappings are based on such matrices, composed, for example, of citations or co-words. Scientometricians have studied the interaction terms mainly in order to position and to rank-order agency in the network, for example in terms of performance. Rank-ordering can be attributed on the basis of relations (e.g., single linkage clustering), but positioning requires study of the structure of the matrix in terms of so-called principal components or eigenvectors. Structure is a property of the matrix that cannot be reduced to the relations among agents (Burt[5]). One can reconstruct it algorithmically, for example by using various forms of factor analysis and multi-dimensional scaling, but one is not able to observe it directly. Structure in a communication network remains fundamentally a hypothesis.

An expectation of structure can be based on available theoretical knowledge, such as that the sciences develop in terms of specialties. Such a theoretically informed framework allows us to appreciate the algorithmically produced results of the scientometric analysis, and for example to recognize specialties in the scientometric mappings. But what is the nature of these hypothesized systems of communication? For example, what makes journals cluster so selectively? Which underlying processes bind together selections by editors, referees, authors, and respective audiences?

Sometimes the analyst is able to point to a clear paradigm like a set of axiomata as a codification, but more often the relevant distributions contain a differentiated structure (cf. Pinch's [6] "evidential contexts"; Amsterdamska & Leydesdorff[7]). A distribution, however, cannot be identified unequivocally as a unit of analysis because it contains an uncertainty. This uncertainty can be expressed as the expected information content of the communicative operation. Note that the definitions of expectation, uncertainty, and theory-based hypotheses are semantically related: the various mechanisms are only indicated by the measurement.

In summary, the skewness in the variation shows that the actors communicate in terms of specific selections. Over time, the selections function as a codification that guide the reproduction of the distributed communication systems. The communication systems are not directly observable, but one is able to specify them as hypotheses. The hypothetical structures are operationalized as distributions that contain an expectation. Without further theorizing one observes only the co-variation between the represented (e.g. cited) and the representing (e.g. citing) systems.

**Regimes of cultural evolution**

A *regime* can be defined as a pattern of communications that is reproduced from year to year. The pattern, however, contains a distribution, and therefore it cannot be identified without uncertainty. A paradigm is then, for example, a special case where the regime is theoretically specifiable. In other cases, the analyst is able to perceive regimes only in terms of their "*instantiations*" at each moment in time (Giddens[8]), and/or in terms of specific "*trajectories*" of densities in the phase space of possible distributions over time (Dosi[9]).

Like paradigms, regimes are contingent in time and space. While trajectories of clusters can be depicted using geometrical metaphors, the regime develops by selecting among possible trajectories in a hyper-space. The discursive analysis, however, can only appreciate a regime by taking a specific perspective, i.e. by focusing on a subdynamic of the system. For example, one is able to analyze the system's construction over time in terms of "variation and stabilization" (e.g., Callon & Latour[10]) or its functioning at each moment in terms of "variation and selection" (cf. Nelson & Winter[11]). Note that orthogonal projections are expected to provide us with nearly incommensurable interpretations of the systems under study (Blauwhof[12]).  The confusion on the qualitative side of the field is thus predictable: formal methods are needed to distinguish the underlying subdynamics.

In biology, one would call the dynamics under study "phenotypical", while the analysis is oriented towards the specification of the composing "genotypes" or subdynamics of the complex phenotype.  The phenotype is complex, since it also contains interaction terms.  However, the "phenotype" of a social system is not identifiable because it remains necessarily distributed, and thus uncertain.  Additionally, a social system contains uncertainty over the time dimension: it is emerging, and thus, it can only be specified as a hypothesis.  The hypotheses provide the social scientist with heuristic guidance, but one should not reify the systems under study.

For example, one can distinguish "genotypically" different function systems in society, such as the economy, government, the sciences, family relations, etc. (cf. Luhmann[13]).  But the delineations among the functions in a distributed system are not given as in a biological body.  Communication systems can "interpenetrate" one another.  Social communications can also be institutionalized, but they need not be: communications can occur both within institutions and across institutional borders.  The institutions can be considered as the fingerprints of the communications that have been useful for the system's reproduction hitherto.  Reproduction means in this context reconstruction, and therefore selection from the possible recombinations along trajectories.  Thus, historical realization is one of the subdynamics of the evolutionary system.  It operates in a distributed mode: the system could also have been different.  There is always a range of possible reproductions of the social system.

Realization of a specific variation indicates specificity with reference to the next-order system.  Specificity is a result of the selectivity of an emerging system.  As noted, variation and selection are structurally coupled like the rows and columns of a matrix.  They are two subdynamics of the same system.  Analogously, stabilization is a subdynamic, i.e., some selections will prove to be stable over time.  Some stabilizations can be selected for globalization.  A regime is then a global super-system that remains virtual by definition, while a trajectory and an instantiation can be depicted using a geometrical representation.

A regime changes continuously as selections operate upon each other in search of (functional) optimalization.  Stabilization is brought about when the selections resonate in a local optimum.  The lower-level systems can adapt to selection pressure by changing their structures as in a drift. Thus, the higher-level system operates as a virtual integrator. As noted, the locus of integration cannot be identified; it remains distributed (cf. Gibbons *et al.*[14]). Thus, the integration reproduces the differentiation that is expected to be functional for the further operation of the system and its subsystems.

All (provisionally codified) structure is continuously under pressure to select from all the perturbances that are generated in lower-level interactions. Functional differentiation is functional for the reproduction of the super-system, but there is a trade-off. The longer a regime prevails, the more it will become differentiated, and therefore operate "on the edge of chaos." If integration between functionally differentiated systems tends to fail, the system suffers from crises. Integration, however, presupposes inter-system communication, i.e. translation of substantive information from one (sub-)system into another. While a trajectory and a scientometric map represent systems of communications, a next-order regime is reproduced in terms of a *system of translations* among communication systems.

## Translation Systems

Translation systems are the carriers of integration. Translations communicate between differentiated systems of communication. Note that communications are less complex than translations: in translations the meaning of a communication is changed.

If the system is functionally differentiated, the relation between the substance of a communication and its meaning can be codified. Scientific communications, for example, are expected to have both an intrinsic substance ("context of discovery") and a function with reference to the next-order system ("context of justification"). Communications contain two degrees of freedom.

Translation systems are additionally able to change the meaning of information. Thus, they are expected to contain three-dimensional informations: in the case of a translation one is able to distinguish the substance, function, and context of the communicated information. While scientific communications can be considered as "universally" true when the context is disregarded, truth tends to become a disciplinary and specific heuristics for puzzle-solving in translation systems.

Each translation is a specific and local integration; the global regime–containing the distribution of translations–remains an analytical construct. Local integration requires a specific perspective, and thus the *position* of a functional subsystem. Thus, the subsystems are nested into one another: translations are not possible without differentiation, and differentiation cannot be sustained without translation. Functions can be stabilized temporarily, for example by codification in institutions. Additionally, functionally differentiated communications can have a value for other subsystems, i.e., in terms of other (but also specific) communications. Note that a range of other subsystems can be functional, and that the translations into the respective codifications is expected to vary accordingly.

One may wish to think of the complex dynamic system as a trajectory in a four-dimensional space. At each location along the trajectory, there is always variation, selection, and stabilization. In general, a system of communications pertaining to a regime of translations is expected to contain these three kinds of dynamics: (i) substantive communication, (ii) recurrence on internal codification (e.g., for quality control), and (iii) output to other subsystems (Luhmann[13]). As noted, a system in a self-organizing regime is able to use its fourth degree of freedom to reshape itself in terms of these combinations.

Unlike a double helix, this triple helix cannot be stable; it is dynamic and even chaotic by nature. It may go through phase transitions, irreversibilities, etc., and it may exhibit all the other well-known species of chaotic behaviour. A four-dimensional regime is sufficiently complex to explain these phenomena (cf. Leydesdorff[15]).

**The hypothetical status of codification**

Selection pressures force the communication systems to stabilize (provisionally) their substance with reference to functions. If repeated over time, the codification can sometimes be institutionalized. But one has to discern analytically between codification and institutionalization. Codification is a structural function at each moment in time; institutionalization is an element of structure in the network that has a function for the retention of structure over time. Codification is reproduced in the communicative operation; institutes may be abolished and/or replaced.

A difficult step in understanding the evolutionary model is the assumption of codification along an axis which remains internal to the (sub-)system. Not only does the variation change, but the structural selection at each moment in time also varies over time. The two layers operate on each other. Even if there is no observable institutionalization of the selective mechanisms, one has to assume codification in a functionally differentiated complex. Without codification a communication system is not able to recur on its previous state. Thus, it would not be able to adapt and henceforth to survive as a (sub-)system.

Since codifications can be internal to the function systems, they may not be directly observable. But the hypothesis of codification can be confirmed on the basis of observables. For example, word occurrences can be provided with meaning, and tacit knowledge can be recognized. This second dimension of the system ("codification") contains a negative (i.e., selective) feedback or an equilibrium function that tends to remain latent or virtual. Its substance has to be hypothesized on theoretical grounds, and structure has accordingly to be inferred from the measurement.

Since codification is based on recursion over time, it introduces a frequency into the system. However, harmony can no longer be assumed *a priori* among the updates in different systems (cf. Latour[16]). The resulting order in the integrating interactions is a consequence of specific adjustments in the mutual and recurrent selections. Systems drift into local optima. Social order is a consequence of the filtering of noise in the communication between otherwise asynchronous systems of communication (Leydesdorff[17]).

In principle, the emerging order is an unintended outcome that develops beyond the control of the participating agents. The "phenotype" looks different from each perspective. Since no single integrating instance is given, the various subsystems have to compete over time to establish their own order in terms of their respective codification. By doing so, they drive one another into (nearly) orthogonal positions. By providing *different* views, more complexity can be handled at the system's level.

The competition of these perspectives is not limited to the reflexive windows. The differences are also codified, and thus the functional differentiation in all systems under study spans a space of possible interactions. Integration is achieved on the basis of a series of translations, but there is no single translator. In short: variation tends to drift into differentiation; differentiation tends to drift into functionality; and functionality in the differentiation drifts towards self-organized criticality because it endangers the system's integration. In a fully differentiated system, each subsystem can claim priority from its own perspective. For example, one can raise questions like whether the market is decisive in controlling society or the political (sub-)system. In the long run is the science/technology system not driving the course of modern societies? Different time axes are involved. The discourses "observe" each other reflexively, in a distributed mode.

Thus, control itself remains in flux and dependent on the selected perspective. Control entails a prediction on the basis of an integration over time from a hindsight perspective. Each subsystem can claim control by projecting inter-system communications onto its internal codification as an axis for the measurement. There is no superior vantage point. One can only look at the system through a reflexive window. Improvement of one's view is perhaps possible by cleaning one's window. We return to the problem of the quality of the reflection in a later section.

## Recursivity of the interactive operation

Systems which co-vary like the rows and columns of a matrix, are structurally coupled. However, communications over the network are no longer coupled structurally, but operationally, since there are then two interfaces: one between the sender and the network, and one beteen the network and the receiver. Thus, there is an additional filter, and consequently, the frequency of an interaction over the network is expected to be an order lower than the communications within each system. For example, while price movements can be extremely fast, the development of capital requires a longer time span, and the development of scientific theories may require decades.

Inter-system communications that involve more than a single selection can be considered as *translations* since a change in the code of the communication has to be assumed. The same transfer of information can be considered as a communication from one vantage point (e.g., the network), while it implies a translation from another point of view (e.g., the receiver of the communication). Thus, the specification of the systems of reference is crucial for achieving analytical clarity. Note that the assessment and impact of a translation is expected to be asymmetrical when different receiving systems are considered as the systems of reference.

The resulting picture of the social system is one of a complex dynamic system of nested translations: the underlying communication systems are relatively high-frequency, and can therefore be considered as constants in a first approximation (Simon[18]). The axes are not fixed, but the communications spin around them while developing historically under the pressures of cultural evolution. The network-system is a next-order system that tends to stand orthogonally with respect to each of the participating systems: it is based on their interactions, and not on their aggregates.

The axes under study turn ninety degrees at each subsequent interface, and the categories have to be specified with reference to each perspective because the orthogonality is expected to lead to incommensurability in the understanding. A receiver is sometimes able to reconstruct a message sent through the network; but the network is by its nature different from both the sender and the receiver that use it. Networks network and actors act: the two operations are structurally coupled in their co-variation, but the systems of reference, i.e., the remaining variances, are different.

The evolutionary hypothesis of near orthogonality among the functional dimensions (Simon[18]) has methodological implications. For example, scientific communication systems should not be considered as an aggregate effect of (groups of) scientific researchers, but as an aggregate of their scientific interactions. Thus, the study of processes of scientific reproduction should not be designed as a (relational) multi-level problem (like the distinction between "groups" and "fields"), but as a problem of "unintended consequences" in a multi-dimensional space. The next-higher level rests as a hyper-cyclic network by selecting from the lower-level ones to which it remains structurally coupled: it "entrains" their development by exerting selection pressure (cf. Kampmann *et al.*[19]). A system is not determined by its contexts, except in the co-variation. For the remaining variation it is conditioned only by contexts (cf. Leydesdorff[20]). For evolutionary reasons, the co-variation is expected to be relatively small (Simon[18]).

In summary, the complex dynamics of the "phenotype" is observable only through the window of a representation, and thus its study can be controlled only from a reflexive stance. The analyst observes variations in the interactions among systems at the lower level, while the selecting structures tend to remain latent. The quality of the view is based on the codification of previous experiences into theories that allow for reconstruction in terms of hypotheses. The quality of the reflexive theories in turn is based on the specificity of these hypotheses and on their methodological selectivity ("rigour") in relation to empirical information. Reflexivity in one's position determines the effectiveness of the possible actions and non-actions.

## Science studies as interacting communication systems

To achieve integration between different windows, for example, between qualitative theorizing and scientometrics, one may wish to choose one hierarchical vantage point or another. This implies a normative decision that sacrifices explanatory power. Integration in terms of mutual translations can only be achieved reflexively. Reflexivity in the translation can provide us with a tool for developing a discursive appreciation of what one can see through specific windows of reflection.

Reflexivity has been widely accepted, both on the more qualitative side of the field of STS and in scientometrics. For example, scientometric information is consciously defined as only an indicator, i.e., a representation of the communication systems under study. The extension of this reflexivity to the translation between different branches of STS may be functional for the improvement of the translation, and thus stimulate the further development of the different areas in terms of their internal codification and mutual integration.

How is one able to study the dynamics of Science & Technology Studies as a differentiated field of science? From an evolutionary perspective, one would expect the emergence of the following subdynamic perspectives in an increasingly differentiated system:

1. The *construction* and emerging stabilization of new structures by interactions among lower-level units;
2. The *use* of knowledge contents and expertise in other subsystems;
3. The *reproduction and modification* of structures which contain codification.

Hitherto, scientometrics has had the programmatic ambition to focus on the latter perspective of "mapping the structure of science" in order to understand the dynamics of the systems under study (e.g., Elkana *et al.*[21]; Small *et al.*[22]). From the second perspective, scientometric results can be used as legitimation for S&T policy decisions. With reference to the sciences and technologies under study, however, this perspective ("utilization") has been the focus of fields like R&D management in private corporations and Technology Assessment in the public arena. These assessments are in need of indicators which remain functional in developing their perspective. Therefore, we witness a continuous effort to extend scientometrics with patent statistics and social indicators. Thirdly, in the sociology of scientific knowledge the perspective of the construction of the systems under study has been programmatic.

In order to provide our reflections with an empirical basis, we performed a citation analysis of major journals of science studies itself. The citation relations among the clusters which could be discerned were further analyzed. We used the following journals as *ego* in the construction of citation networks during the period 1980-1994: *Scientometrics*, *Social Studies of Science*, and *Research Policy*. The citation patterns of *Science, Technology, & Human Values* were also analyzed, but this journal changed in character in the middle of the period (1988) when it became the journal of the *Society for the Social Studies of Science (4S)*.

The methods which we used are analogous to the ones which we detailed in Van den Besselaar & Leydesdorff[23] for the reconstruction of the development of Artificial Intelligence. We fully analyzed the citation environments and patterns for each of the even years (1980, 1982, etc.), both in the cited and in the citing dimension, and using a one percent threshold.[1] We intend to report on the empirical findings of this research in another article (Van den Besselaar & Leydesdorff, in preparation),[2] but we draw here below on the conclusions from these analyses in relation to the theoretical argument about communication systems as it has been developed above.

As noted, the assumption is that *Social Studies of Science* can be considered as a codifier of substantive communication, while *Scientometrics* mainly codifies formal communication in this area. *Research Policy* can be considered as a journal at the interface that draws on science studies both formally and substantively. *Science, Technology & Human Values* has a programmatic title that indicates an integrating role, but the expectation is that in practice the selection in this journal is biased toward substance because of its constituency.

The analysis of the citation relations between these STS journals reveals that the citation networks of *Scientometrics* and *Social Studies of Science* have grown apart during the 1980s. Taking *Scientometrics* as the ego, these two journals were part of a single cluster in 1980 (*Figure One*). In the equivalent network for 1994 (*Figure Two*), the two groupings are separated. The analysis of in-between years shows that the years with intensive citation traffic between the two clusters are scarce (cf. Leydesdorff[24]). The pattern of mutual exchange between the two core journals is declining over the years (*Figure Three*).

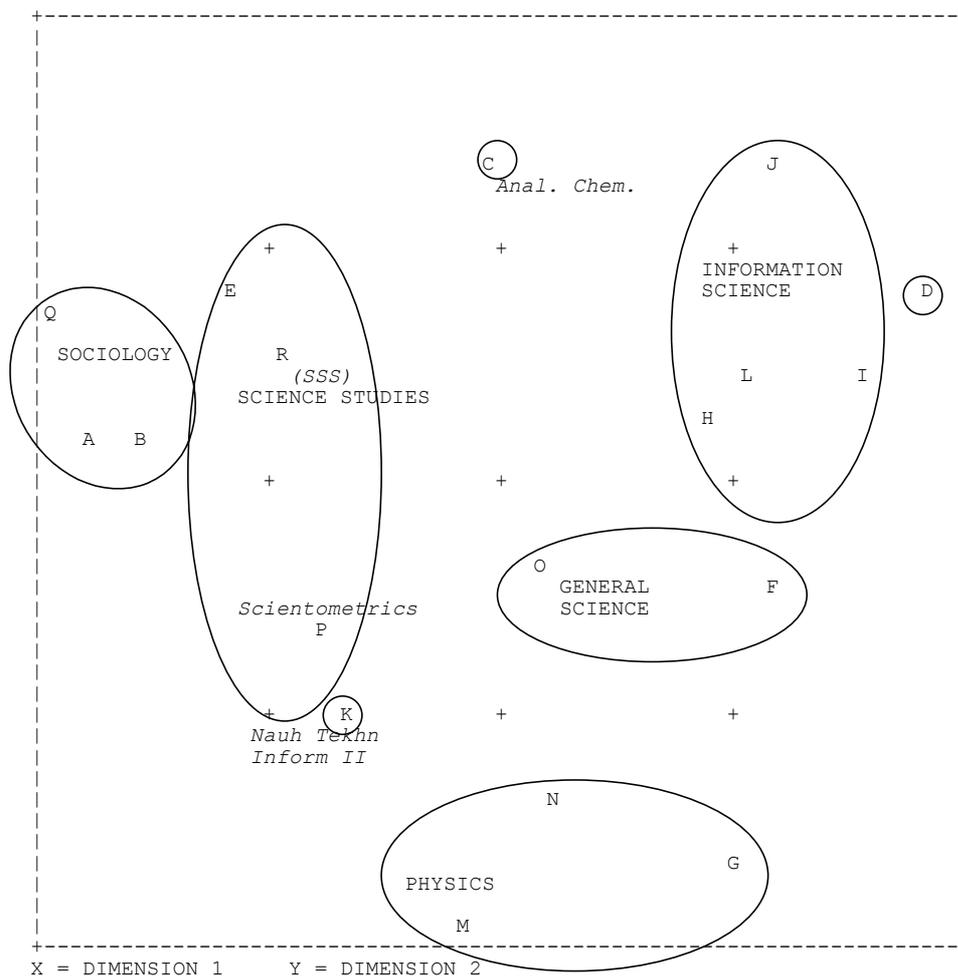

**Figure 1**
Plot of Stimulus Space based on Factor-analysis and MD-SCAL for
*Scientometrics* as *ego* in its 1980 citation environment
(citing patterns; threshold = 1%)

Journal name abbreviation

| | | | |
|---|---|---|---|
| A. | AM J SOCIOL | P. | SCIENTOMETRICS |
| B. | AM SOCIOL REV | Q. | SIMULATION GAMES |
| C. | ANAL CHEM | R. | SOCIAL STUD SCI |
| D. | ASLIB PROC | | |
| E. | CURR SOCIOL | | |
| F. | CURR CONT | | |
| G. | CZECH J PHYSICS | | |
| H. | JASIS | | factor designation |
| I. | J DOC | | |
| J. | NACHR DOK | I. | Science Studies |
| K. | NAUCH TECHN INFORM II | II. | Information & Library Science |
| L. | PASIS | III. | Sociology |
| M. | PHYS REV D | IV. | Physics |
| N. | PHYS REV LETT | 12 V. | General Science |
| O. | SCIENCE | | |

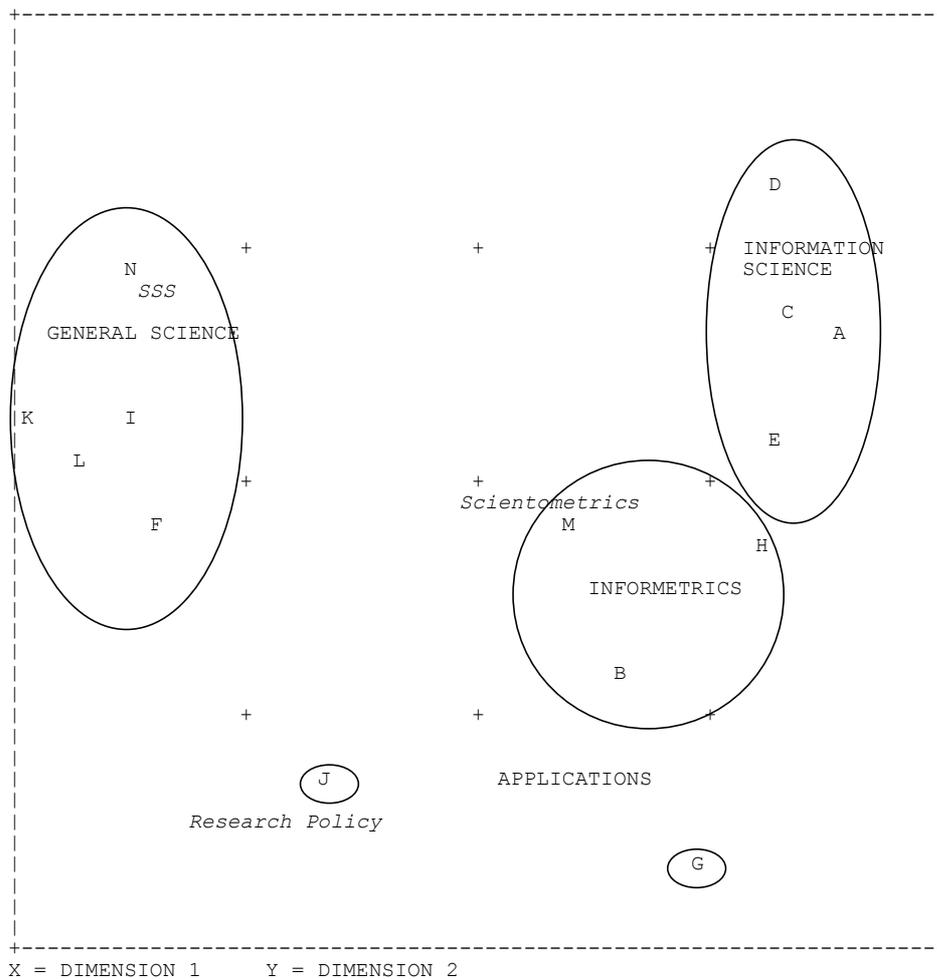

**Figure 2**
Plot of Stimulus Space based on Factor-analysis and MD-SCAL for
*Scientometrics* as *ego* in its 1994 citation environment

```
Journal name abbreviation                    factor designation

A.   INFORM PROCESS MANAG                    I.    General Science
B.   INT FORUM INFORM DOC                          (incl. SSS, ST&HV)
C.   J AM SOC INFORM SCI                     II.   Information Science
D.   J DOC                                   III.  Informetrics
E.   J INFORM SCI                            IV.   Applications
F.   J SCI IND RES INDIA
G.   LIBR ACQUIS PRACT TH
H.   NACHR DOK
I.   NATURE
J.   RES POLICY
K.   SCI TECHNOL HUM VAL
L.   SCIENCE
M.   SCIENTOMETRICS
N.   SOC STUD SCI
```

(citing patterns; threshold = 1%)

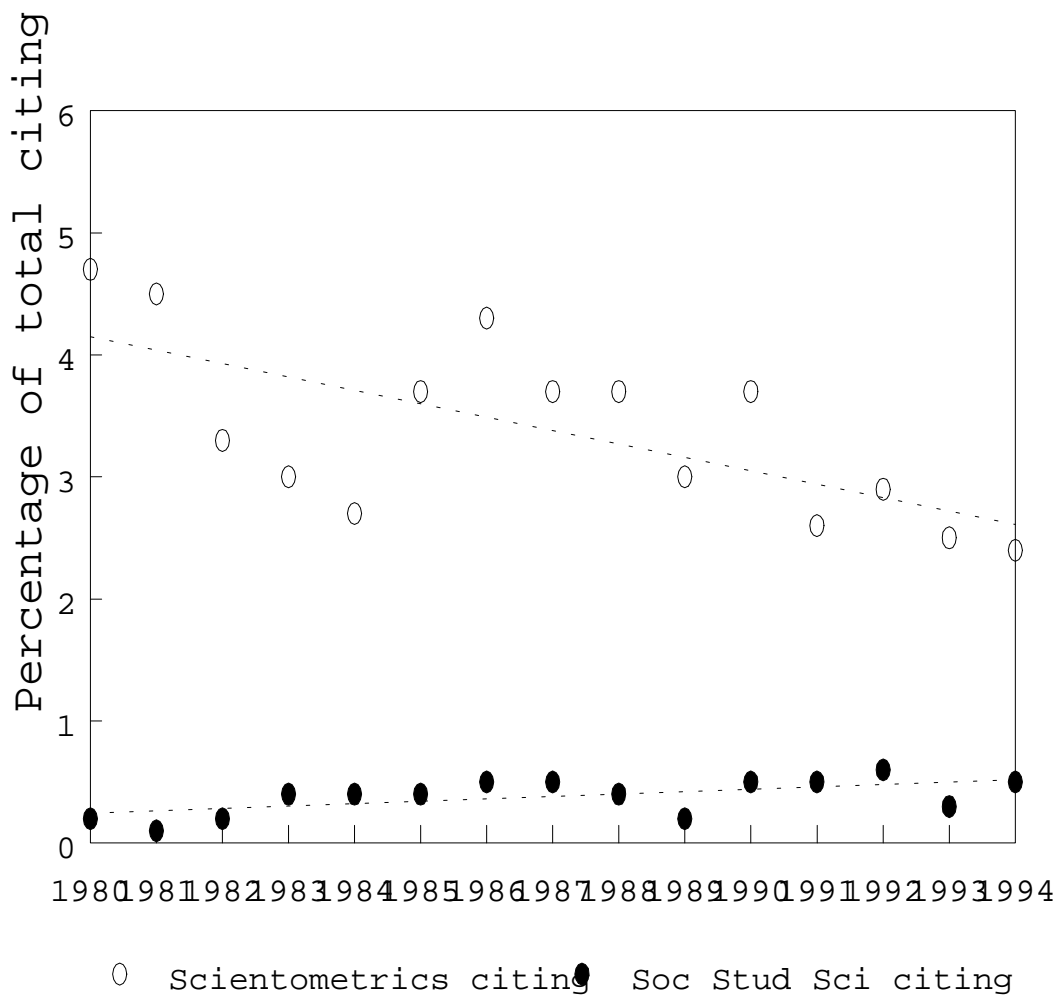

**Figure 3**. Three year moving averages of citation relations between *Scientometrics* and *Social Studies of Science*.

This last figure also shows that the citation relation between the journals *Scientometrics* and *Social Studies of Science* is asymmetrical: while articles in *Scientometrics* cite papers from *Social Studies of Science* on a regular basis, the articles in *Social Studies of Science* which contain references to studies in *Scientometrics* are specific. In general, *Scientometrics* is grouped among a set of "information science" journals in the citation environment of *Social Studies of Science*. *Social Studies of Science* is used as a source journal in the citation environment of *Scientometrics*, together with journals like *Science* and *Nature*, but at a lower level.

*Research Policy* is not present in the citation environment of *Scientometrics* in 1980 nor 1982. *Scientometrics*, however, is always present in the citation environment of *Research Policy*. The mutual citation relations between these two journals become firm and stable from 1984 onwards, as research evaluation studies begin to pay systematic attention to bibliometric indicators (Martin & Irvine (1983)[25]; Moed *et al*. (1985)[26]).[3]

*Research Policy* is present in the citation environment of *Social Studies of Science* in 1980, 1984, and 1986.  In this last year, it is analyzed as belonging even to the same cluster as *Scientometrics* in this environment.  After 1988 *Social Studies of Science* is also no longer present in the citation environment of *Research Policy*.  Thus, citation relations between these two journals are incidental in 1990, 1992, and 1994.

As noted, *ST&HV* should be considered in this context only for the period after 1988.  Although it has a different origin, this journal increasingly joins *SSS* in a single cluster, both in terms of its own citation patterns, and in its position in the citation environment of *Scientometrics*.[4]  Citations to *Scientometrics* from these two journals are often based on (co-)authorship relations among documents with authors who publish in *Scientometrics* as their major publication outlet.

## Differentiation in Science & Technology Studies?

What do these findings teach us about the differentiation of communication structures in STS as a reflexive specialty?  First, *Social Studies of Science* can be considered as a source of citations.  In citation analysis, one is used to hierarchical classifications of "cited" as source, and "citing" as sink.  However, from the perspective of the inter-journal communications, the hierarchical ranks which can be attributed to the journals as *communicating* structures are not at issue.  Given our research question, the "reception" of communication from the perspective of hindsight is crucial.

From this perspective, "citing" in the present becomes more important for further development than being "cited" because of past performance.  (The inversion illustrates that the categories have to be specified anew when one changes the system of reference.[15; 28])  Our results indicate that the relation between qualitative theorizing and scientometric methods has stabilized over the last decade with the qualitative side being locked into the position of providing source materials for the quantitative side.  This can formally be considered as an achievement.  However, from our substantive reading of the journals, we have hitherto not been aware of a systematic use of qualitative insights into scientometric modelling. Perhaps a necessary but not a sufficient condition for further development is being fulfilled. We shall return to this speculation in a later section.

Second, the differentiation between the "utilization" axis and the formal axis also seems strongly developed if one considers *Research Policy* as an operationalization of the former. Neither *Scientometrics* nor *Research Policy* places itself in the same citation cluster as the other in any of the years analyzed. In the later years, *Social Studies of Science* is no longer present in the citation environment of *Research Policy* (using a threshold), while in earlier years it had appeared in the same cluster as *Scientometrics*. Thus, the relations with the substance of STS seem to be formalized.

In summary, the perspective of functions for the communication network has provided us with a theoretically informed, but different interpretation of cited/citing relations and their structure. Hitherto, the focus in scientometrics has been on hierarchical ranking of *cited* papers (or authors), for example in terms of "quality". While communicating agents can be ranked, it makes little sense to rank functions of the communication system. Functions among differentiated systems are expected to be different; but these different functions are needed for the integration and reproduction of the system. Functional analysis cannot be understood in terms of hierarchies; functions tend to be orthogonal to one another, and therefore they will eventually be heterarchical.

## Communication theoretical perpsectives

Integration is based on local translations. Thus, one may expect integration within each of the perspectives, capitalizing on the strength of their own axes. Can the expected nature of these different translations be specified? Is one able to make comparisons among them? Let us proceed by specifying the expectation for integration along the various axes of the STS system.

First, the focus of integration along the "utilization" axis (*Research Policy*) is expected to be found in the *other* system represented in this translation, since this is the direction of the knowledge flow. The substance of the communication is selected from a user's perspective.

Second, one expects actor-centered integration along the axis of substantive variety which predominates in the sociology of scientific knowledge. Each actor can provide the observed mechanisms reflexively with meaning. The variety of meanings can, of course, be made the subject of formal analysis, but it also informs a reflexive actor. From the constructivist perspective, however, this information is structured with reference to "enabling and constraining" conditions for further action, and not with reference to the "unintended outcomes" of interaction at the level of communication between the agencies involved (although this problem has reflexively been recognized; cf. Giddens[8]; Beck[29]).

What type of integration may we expect on the formal side of STS? Is the formal integration of the structure of the relations between the different reflections an option? Paradoxically, such an integration would require a theoretical reflection. But since this is also a formal reflection, the theorizing becomes mathematical. For example, one can raise the formal question of how many angles are possible, if each of the perspectives chooses a specific angle to study the system. Of course, an infinite number of angles is possible; but along how many angles does one expect codification? As noted, functions are expected to develop orthogonally, and therefore this question can be reduced to the simpler question of how many *orthogonal* representations one could expect.

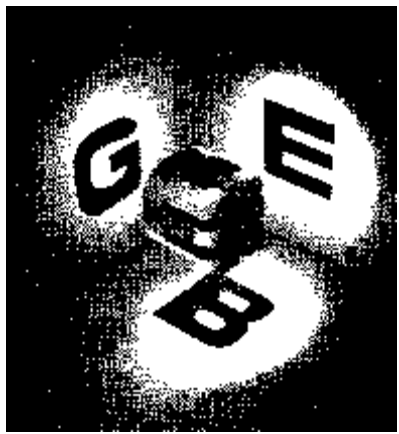

**Figure 4**

Hofstadter's (1979) triplet *Gödel-Escher-Bach*

*Figure Four* shows Hofstadter's[30] well-known Triplet *Escher-Gödel-Bach*. A three-dimensional object has three orthogonal projections in two-dimensional planes. In general, an N-dimensional object has N orthogonal projections in N-1 dimensional spaces. Thus, a changing object has 4 orthogonal projections in three-dimensional geometries. Each of these geometries has a blind spot: using a geometrical metaphor a perspective is necessarily taken (cf. Luhmann[13]). For the case of STS, we have hitherto specified only three of these perspectives. The fourth perspective is again formal, since it is based on this mathematical reflection. It has a blind spot for the substantive variation.

In which respect would this formal perspective differ from the ones which we have already specified above? The other subdynamics kept the variety in perspective by using respectively the axis of (a) variety and stabilization, (b) variety and utilization, and (c) variety and structural codification. Since this reflection on the other reflections lacks empirical substance, it can be inferred only on the basis of these other analyses, and it cannot be perceived in the data. But its development may allow us to integrate the three noted perspectives into a formal model.

From the perspective of this model, the insights of the three other perspectives are selective conditions which have operated in a space of possible combinations. Thus, the model considers the phase space and translates the positive insights of the substantive perspectives on the events which do happen into selective, i.e. negative, operations on the wealth of events that might have happened.

In principle, this operation can be performed by using an algorithmic computer language. The code translates the substantive insights of the other perspectives into selective conditions or "do while" loops that can search the phase space for possible combinations. On the basis of the "genotypical" specifications by substantive theorizing, the model can thus bootstrap into exploring other possible "phenotypes" of the system.

The results of such a model can be appreciated from each of the lower-level perspectives. As noted above, the underlying perspectives are expected to compete in terms of understanding the results of the model system. The model can thus be considered as an integrating machinery among the insights along different axes. Note that this function of modelling is well known in other disciplines, e.g., in economics and cognitive science (e.g., Rumelhardt et al.[31]). At this moment, it is only programmatic in STS (cf. Andersen[32]; Leydesdorff[20]).

In summary, a dual function can be specified for the formal approach. First, the development and production of scientometric indicators informs us about the codified dimension of the communication systems under study, and second, the development of algorithmic models may allow us to import insights from each of the underlying perspectives.

While scientometric indicators have hitherto focused on the analysis of available data, the modelling effort is based on reflexive differentiation of the theoretical perspectives. The differences among the codes can be codified by using computer language. The specifications can then be used for bootstrapping to the phase space of possible specifications.

## Policy implications

As noted, one expects the utility function to be different from the intrinsic codification, and thus to be indicated by other indicators. The utility of communications that cross system boundaries is determined by the receiving system and in terms of the latter's codification. As the utility function develops, it can be codified with hindsight.

The sciences functioned mainly to rationalize subsystems of society during the Enlightenment of the 18th century. The modern industrial system exploits the sciences in terms of potential innovations. Patents have been considered as potential indicators of this relation. While initially emergent in the interaction between the sciences and the economy, the patent system has perhaps itself become a reflexive communication system.

Since patents are codified, they can be counted, and thus in scientometrics the question emerged of whether one should focus on patents or on publications. Narin and Olivastro[33], for example, have argued that the patent system tends to fuse with the science system in the case of biotechnology. Similar claims about a shift of focus from the traditional scientific communication system to "the network level" have been made on sociological grounds by Gibbons *et al.*[14], and more recently by Katz *et al.*[34] on the basis of scientometric data for the UK. Blauwhof[35], however, found considerable differentiation of patents and scientific publications in the case of telecommunication. Thus, the cycles of interactions may be technology-specific, and both the structural level which generates the variation and the one which operates selectively may change over time (cf. Barras[36]).

When does the emerging interaction system take control? In general, when a translation system is codified *between* two differentiated systems, a triple helix is generated and complex dynamic processes can be expected to emerge (cf. Leydesdorff & Etzkowitz[37]). Complex data, however, mean that we are no longer able to specify the relevant subdynamics without theoretical assumptions. Various interpretations may be equally valid.

Combining (potentially multiple) interpretations into a reflexive model may provide us with a foothold for a more systematic understanding.

[return](#)

**Notes**

1. Citation matrices are constructed on the basis of an *ego* (journal), including all journals citing from or cited by this journal at one percent of its total citation rate in the respective dimension. Matrices are factor analyzed (using VARIMAX rotation), and multi-dimensional scaling of the correlation matrices (MINISSA) is used for visual representation of the results.

2. This paper was later published as P. Van den Besselaar (2001) The cognitive and the social structure of Science and Technology Studies. *Scientometrics* 51, 441-460.

3. Before the publication of Martin & Irvine (1983),[25] *Research Policy* had published incidental papers using scientometric methods (e.g., Chang & Dieks, 1976[27]).

4. *ST&HV*, however, was present in the citation environment of *Research Policy* in 1994.

[return](#)